# Statistical simulation of pattern formation on plane target surfaces by ion beam sputtering

Nikolay A. Kudryashov[*], Mikhail V. Skachkov

*National Research Nuclear University MEPhI (Moscow Engineering Physics Institute), 31 Kashirskoe sh., 115409 Moscow, Russia*

_______________________________________________________________________

**Abstract**

Numerical simulation of pattern formation on plane target surfaces undergoing ion-beam sputtering is carried out. Base of the mathematical model of target ion-sputtering is nonlinear evolutionary equation in which the erosion velocity dependence on ion flux is evaluated by means of a Monte Carlo method. This approach needs much computational resources, but allows to investigate the influence of ion flux parameters on surface topography. Deviation of the findings from the pattern formation predicted by the continuum model is discussed.

_______________________________________________________________________

## 1. Introduction

Since the 1970s research of the nature of ripple pattern forming on metal and semiconductor surfaces at ion-sputtering has been carried out. Numerous experimental studies on monoenergetic ion-beam normal incidence on plane target surfaces have demonstrated ordered structures from hillocks or depressions with hexagonal symmetry formation. At ion-beam off-normal incidence a ripple pattern has appeared on the surface. The orientation of the ripples depends on the angle of the ion-beam incidence. The wave vectors of the ripples are parallel to the projection of the ion-beam direction on the plane target surface for small incidence angles and perpendicular to the projection for gazing incidence. The wavelength of the ripples is usually of the order of tenths of micrometers. This phenomenon has found practical use in nanoelectronics and nanophotonics. Detailed review of experimental works and examples of practical use of periodic structures on metal and solid surface can be found in the works [1-2].

To theoretically explain the ripple forming the continuum model is used. The model basis is the following equation [2]:

$$\frac{\partial h}{\partial t} = -V_0 \sqrt{g} - K \cdot \nabla^4 h + \eta(x, y, t),  \tag{1}$$

_______________________

[*]Corresponding author.
*E-mail address:* nakudr@gmail.com (N. A. Kudryashov).

where $h(x,y,t)$ is the surface height at time $t$ above point $(x, y)$ on the surface plane; $V_0$ is the normal erosion velocity; $g = 1 + (\partial_x h)^2 + (\partial_y h)^2$; $K$ is the relaxation rate due to surface diffusion; $\eta(x,y,t)$ is a noise term describing the random nature of incoming ions and is a set of uncorrelated random numbers with zero configurational average. The normal erosion velocity is proportional to total power deposited per unit volume at the point $(x, y, h(x,y,t))$:

$$V_0 = -\Lambda \iint \mathbf{J} \cdot \mathbf{n}' \cdot E(x,y,h(x,y,t),x',y',h(x',y',t)) \cdot dS', \tag{2}$$

where $\mathbf{J}$ is the average ion flux; $\mathbf{n}'$ is the outward-pointing normal to the surface at the point $(x',y',h(x',y',t))$; $E(x,y,h(x,y,t),x',y',h(x',y',t))$ is the energy deposited per unit volume at the point $(x,y,h(x,y,t))$ by an ion striking the surface at the point $(x',y',h(x',y',t))$; $dS'$ is the surface element at the point $(x',y',h(x',y',t))$.

The partial integro-differential equation (1-2) is hard to analyze. In the works [2-7] the authors reduce this equation to the stochastic nonlinear partial differential equation of the fourth or sixth order. At that almost all theories are based on the Sigmund's sputtering theory [8], according to which the ion depositing energy is approximated by the Gaussian distribution. Then authors studied the morphological features that the model predicts and gave the experimental approval of the results obtained. The surface morphology resulting from the theoretical and numerical analysis correlates with experimental observations. On the ion-beam off-normal incidence the wave vectors of the ripples are parallel to the projection of the ion-beam direction for small incidence angles, and perpendicular to the projection for gazing incidence. In case of ion-beam normal incidence on surface the patterns in the form of depressions or hillocks are observed. At that any symmetry does not result from the continuum model.

However, some features of the phenomenon observed in the experiment are not explained by the continuum model. Among these is the ion source parameters influence on structure formation features on ion-sputtered surfaces. Experiments proved that variation of accelerating potential is very important for the structure formation [9]. For some values of this parameter the sputtering surface structure is not quite ordered, for certain others it is ordered well. Authors of the review monograph [1] admit that importance of the ion source individual settings to the structure forming can be the reason why different research groups get conflicting results concerning the structure formation at Si. Therefore, theoretical research of the ion source specific parameters which can influence on the periodic structure formation on solid surface is necessary.

In this paper, numerical simulation of pattern formation on plane target surfaces undergoing ion-beam sputtering is carried out. Base of the mathematical model of target ion-sputtering is nonlinear evolutionary equation, in which the erosion velocity dependence on ion-flux is evaluated by means of a Monte Carlo method. This approach needs much computational resources, but allows investigating the influence of ion-flow parameters on surface topography. Deviation of the findings from the pattern formation predicted by the continuum model is discussed.

## 2. Problem statement

Collimated monoenergetic ion-beam incidence on plane target surfaces is studied. Evolution of the target surface profile in time is described by the following equation:

$$\frac{\partial h}{\partial t} = -V_0 \sqrt{g} - D \cdot \left( \frac{\partial^4 h}{\partial x^4} + \cos^4 \theta \frac{\partial^4 h}{\partial y^4} + 2\cos^2 \theta \frac{\partial^4 h}{\partial x^2 \partial y^2} \right), \tag{3}$$

where $g = 1 + (\partial_x h)^2 + (\partial_y h)^2$,

$$V_0(x, y, t) = \Lambda \iint dx' dy' \cdot \Phi(x', y') \cdot E(x - x', y - y', h(x, y, t) - h(x', y', t) + a), \tag{4}$$

$$E(x, y, z) = \frac{1}{(2\pi)^{3/2} \mu^2 \sigma} \exp \left\{ -\frac{z^2}{2\sigma^2} - \frac{x^2 + y^2}{2\mu^2} \right\},$$

$a$ is an average ion penetration,

$\Phi(x, y)$ is an ion flow density perpendicular to a plane $xy$,

$D$ is a surface diffusion coefficient,

$\theta$ is an angle between the target plane and the plane $xy$ (axis x is parallel to the target plane). At $\theta = 0$ the ion incidence on the plane target is normal.

Parameters $\Lambda, a, \sigma, \mu, D$ are constant, i.e. not dependent on variables $x, y, t$.

Problem statement includes equation (3), which is solved in the rectangular area $0 < x < H_x, 0 < y < H_y$ at $t > 0$, initial condition is:

$$h(x, y, 0) = -y \cdot \operatorname{tg} \theta \tag{5}$$

and periodic boundary conditions are:

$$\begin{aligned} &h(x + H_x, y, t) = h(x, y, t), \\ &h(x, y + H_y, t) = h(x, y, t) - H_y \cdot \operatorname{tg} \theta. \end{aligned} \tag{6}$$

Meanwhile the ion flow density $\Phi(x, y)$ is considered to be a periodical function of $x$ and $y$ variables with periods $H_x$ and $H_y$ respectively.

At $\Phi(x, y) = \Phi_0 = const$ the problem solving (3-6) is the following:

$$h = -\frac{\Lambda \cdot \Phi_0}{\sqrt{2\pi(\sigma^2 \cos^2 \theta + \mu^2 \sin^2 \theta)}} \exp \left\{ -\frac{a^2}{2(\sigma^2 + \mu^2 \operatorname{tg}^2 \theta)} \right\} \cdot t - y \cdot \operatorname{tg} \theta, \tag{7}$$

i.e. the surface stays plane at any given time.

However, it can be noted that at times associated with interaction time of one ion with the target material the ion flux should not be considered as a continuum medium. The ion flux is a set of separate particles falling on the target surface at different times:

$$\Phi(x, y) = \sum_k \delta(x - x_k) \cdot \delta(y - y_k) \cdot \delta(t - t_k),$$  (8)

where $\delta(x)$ - is the Dirac delta function.

Substitution of function (8) into integral (4) results in the following expression for erosion velocity:

$$V_0(x, y, t) = \Lambda \cdot \sum_k E(x - x_k, y - y_k, h(x, y, t) - h(x_k, y_k, t) + a) \cdot \delta(t - t_k)$$  (9)

The resulting equation contains new parameters $x_k, y_k, t_k$, which are naturally considered as position of points equally distributed in rectangular parallelepiped $0 < x < H_x, 0 < y < H_y, 0 < t < T$.

### 3. Numerical solution of problem

The following numerical scheme of solving the problem (3), (5-6), (9) is implemented.

Uniform rectangular grid is put in

$$\Pi = \{i\Delta x, j\Delta y\}, \quad i = 0...N_x, \quad j = 0...N_y, \quad \Delta x = H_x / N_x, \quad \Delta y = H_y / N_y.$$  (10)

Differential equation (3) is approximated by difference equation

$$\frac{h_{ij}^{n+1} - h_{ij}^n}{\tau} = -\tilde{V}_0 \sqrt{\tilde{g}} - D \cdot (\Lambda_x[h_{ij}^n] + \Lambda_y[h_{ij}^n] + 2\Lambda_{xy}[h_{ij}^n]),$$  (11)

where $\tau$ is the time step,

$$\Lambda_x[h_{ij}^n] = \frac{h_{i+2j}^n - 4h_{i+1j}^n + 6h_{ij}^n - 4h_{i-1j}^n + h_{i-2j}^n}{(\Delta x)^4},$$

$$\Lambda_y[h_{ij}^n] = \frac{h_{ij+2}^n - 4h_{ij+1}^n + 6h_{ij}^n - 4h_{ij-1}^n + h_{ij-2}^n}{(\Delta y)^4} \cos^4 \theta,$$

$$\Lambda_{xy}[h_{ij}^n] = \frac{h_{i+1j+1}^n + h_{i-1j-1}^n + h_{i+1j-1}^n + h_{i-1j+1}^n - 2(h_{i+1j}^n + h_{ij+1}^n + h_{i-1j}^n + h_{ij-1}^n) + 4h_{ij}^n}{(\Delta x \Delta y)^2} \cos^2 \theta,$$

$$\tilde{g} = 1 + (\Delta_x h_{ij}^n / \Delta x)^2 + (\Delta_y h_{ij}^n / \Delta y)^2,$$

$$\Delta_x h_{ij}^n = \begin{cases} 0, & \min(h_{i-1j}^n, h_{ij}^n, h_{i+1j}^n) = h_{ij}^n, \\ h_{ij}^n - h_{i-1j}^n, & \min(h_{i-1j}^n, h_{ij}^n, h_{i+1j}^n) = h_{i-1j}^n, \\ h_{i+1j}^n - h_{ij}^n, & \min(h_{i-1j}^n, h_{ij}^n, h_{i+1j}^n) = h_{i+1j}^n, \end{cases}$$

$$\Delta_y h_{ij}^n = \begin{cases} 0, & \min(h_{ij-1}^n, h_{ij}^n, h_{ij+1}^n) = h_{ij}^n, \\ h_{ij}^n - h_{ij-1}^n, & \min(h_{ij-1}^n, h_{ij}^n, h_{ij+1}^n) = h_{ij-1}^n, \\ h_{ij+1}^n - h_{ij}^n, & \min(h_{ij-1}^n, h_{ij}^n, h_{ij+1}^n) = h_{ij+1}^n. \end{cases}$$

$$\tilde{V}_0 = \Lambda \cdot \sum_k E(i\Delta x - x_k, j\Delta y - y_k, \overline{h}_{ij}^n - h_k + a) \cdot (\eta((n+1)\tau - t_k) - \eta(n\tau - t_k)) / \tau, \qquad (12)$$

$\eta(t)$ is the Heaviside function,

$$\overline{h}_{ij}^n = (h_{ij}^n + h_{i+1j}^n + h_{i-1j}^n + h_{ij+1}^n + h_{ij-1}^n) / 5, \qquad (13)$$

$h_k$ is the target surface height in the ion incidence point which value is calculated by bilinear interpolation from the grid nodes (10) on the time layer *n*. Summation in the right part of the formula (12) is actually carried out using all ions falling on the target at times between time layers *n* and (*n*+1).

Difference equation (11) supplemented by initial and boundary conditions (5-6) is an obvious conditionally stable numerical scheme. To provide stability of the scheme the following conditions are to be met:

$$D \cdot \tau \cdot \left[ \frac{1}{(\Delta x)^2} + \frac{\cos^2 \theta}{(\Delta y)^2} \right]^2 < \frac{1}{8}, \qquad (14)$$

$$\frac{\Lambda}{(2\pi)^{3/2} \mu^2 \sigma} < \Delta x / 2, \qquad \frac{\Lambda}{(2\pi)^{3/2} \mu^2 \sigma} < \Delta y / 2, \qquad 2\pi\mu^2 \Phi_0 \tau < 1. \qquad (15)$$

The first inequality (14) is the condition of the difference scheme stability without nonlinear term $\tilde{V}_0 \sqrt{\tilde{g}}$. Inequalities (15) provide evaluation validity

$$\tilde{V}_0 \sqrt{\tilde{g}} \cdot \tau \le \tilde{V}_0 \cdot \tau + \frac{1}{2} |\Delta_x h_{ij}^n| + \frac{1}{2} |\Delta_y h_{ij}^n|.$$

Coordinates $x_k, y_k$ of points of interaction between ions and the target surface make sequence of points uniformly distributed in rectangle $0 < x < H_x, 0 < y < H_y$. According to the Monte Carlo standard method they are uniformly distributed independent random points. Their coordinates are sampled using random number generator which is part of standard libraries of any programming language.

The present numerical scheme also implements the other approach according to which the sequence of independent random points uniformly distributed in $N$-dimensional cube is replaced by their determinate analogue that is Holton sequence [10]. In two-dimensional case the position of points $(p_k, q_k)$ of the Holton sequence is calculated with the following formulas:

$$p_k = \sum_{i=1}^{m} a_i \cdot 2^{-i}, \quad q_k = \sum_{i=1}^{n} b_i \cdot 3^{-i},$$

where $k = a_m a_{m-1}...a_1$ is an index written to dual number system, $k = b_n b_{n-1}...b_1$ is an index written to ternary number system. The first Holton sequence members are: (0, 0), (1/2, 1/3), (1/4, 2/3), (3/4, 1/9), (1/8, 4/9), (5/8, 7/9), (3/8, 2/9), (7/8, 5/9),... .

For any natural number $K$, $s$ and $l$ from $K$ first points of sequence $(H_x p_k, H_y q_k)$ exactly $\left[ \dfrac{K}{N_x N_y} \right]$ or $\left[ \dfrac{K}{N_x N_y} \right]+1$ points fall in any cell $[i\Delta x, (i+1)\Delta x) \times [j\Delta y, (j+1)\Delta y)$ of uniform rectangular grid (10) with parameters $N_x = 2^s$, $N_y = 3^l$. This feature provides the better equitability of the Holton sequence points in comparison with the sequence of independent random points resulting in better integral calculation by the Monte Carlo method. The best equitability of the Holton sequence points is shown in Fig. 1-2. Fig.1 shows $K = 10^4$ first Holton sequence points and Fig.2 shows $K = 10^4$ independent random points uniformly distributed in the unit square.

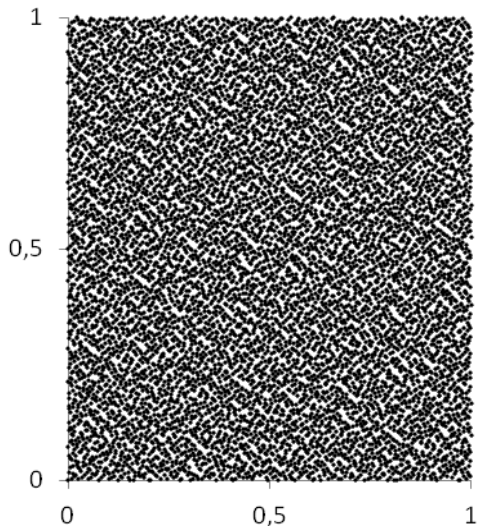

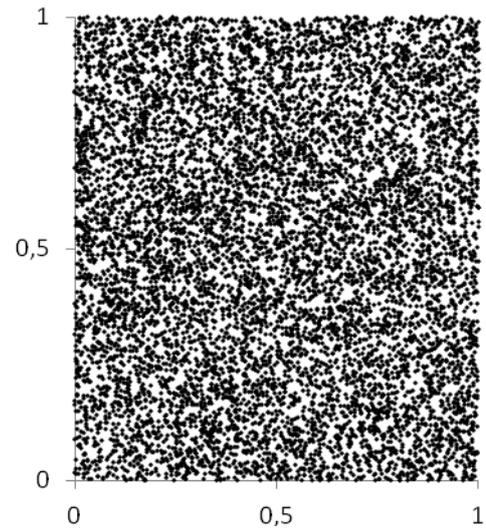

**Fig.1.** $K = 10^4$ first Holton sequence points

**Fig.2.** $K = 10^4$ independent random points uniformly distributed in the unit square

The above explained numerical scheme of the problem solving is implemented using resources of the RAS Interdepartmental Supercomputer Center. Appropriate program code is made for multiprocessor MVS-100k using MPI functions. To parallelize the algorithm the domain decomposition methods with overlapping subdomains is used. The given geometrical method of parallelizing is effective enough for explicit numerical schemes of boundary problem solving. Surface erosion velocity calculation by the Monte Carlo statistical method does not make serious difficulties, because ion contribution to erosion velocity decays with increase of distance from the ion according to the Gaussian law, and summation in the formula (12) can be carried out in the range of quite little neighborhood of computational point.

It should be pointed out that contribution to the target surface erosion velocity in boundary points of computational domain is made by the ions situated beyond it. However, as far as ion flux density $\Phi(x, y)$ is periodic function of variables $x$ and $y$ with periods $H_x$ and $H_y$ respectively and solution meets conditions of periodicity (5-6) simultaneously with each ion situated inside the computational domain the ions situated beyond it are compatible with the target. Their coordinates are also found out from periodicity conditions.

## 4. Calculation results

Basic calculations are made on the grid $N_x \times N_y = 1250 \times 1250$ with time step equal to $\tau = 1250$ at the following parameter values:

$$a = 1, \quad \sigma = \mu = 1/3, \quad \frac{\Lambda}{(2\pi)^{3/2} \mu^2 \sigma} = 0.03, \quad D = 10^{-8}.$$

Interval between two sequential time moments of ion incidence on the target surface in the range of the computational domain is taken as time unit, i.e. time is measured in numbers of ions fallen within the rectangular domain $0 < x < H_x, 0 < y < H_y$ in all previous target surface sputter time. In this connection the formula (8) should show $t_k = k$. The following four variants of sizes of computational domain and angle of target plane with respect to plane $xy$ are considered:

1) $\text{tg}\,\theta = 0, \quad H_x = 200, \quad H_y = 200$ (normal incidence);          (16)

2) $\text{tg}\,\theta = 1/2, \quad H_x = 200, \quad H_y = 175$;          (17)

3) $\text{tg}\,\theta = 1, \quad H_x = 200, \quad H_y = 125$;          (18)

4) $\text{tg}\,\theta = 2, \quad H_x = 200, \quad H_y = 100$.          (19)

The problem solving is shown as follows:

$$h = u(x, y) - h_0(t) - y \cdot \text{tg}\,\theta, \tag{20}$$

where $h_0(t) = -\dfrac{1}{H_x H_y} \displaystyle\int\limits_0^{H_x} \int\limits_0^{H_y} (h(x, y) + y \cdot \operatorname{tg}\theta) dx dy$ - target sputtering depth.

Solution presentation in the form of (20) is easy to compare with solution of problem (7) which is true at $\Phi(x, y) = \Phi_0 = const$.

Figs. 3-10 show the function $u(x,y)$ calculation results, which corresponds to top view (in direction of incident flux) for all four cases (16-19). Black color corresponds to minimum function value, white color − to maximum function value. Figs. 3-6 show the results in case when coordinates of ions in incident flux are coordinates of independent random points uniformly distributed in the unit square (Fig.2). Figs. 7-10 show the similar results in case when coordinates of ions in incident flux are coordinates of the Holton sequence points (Fig.1).

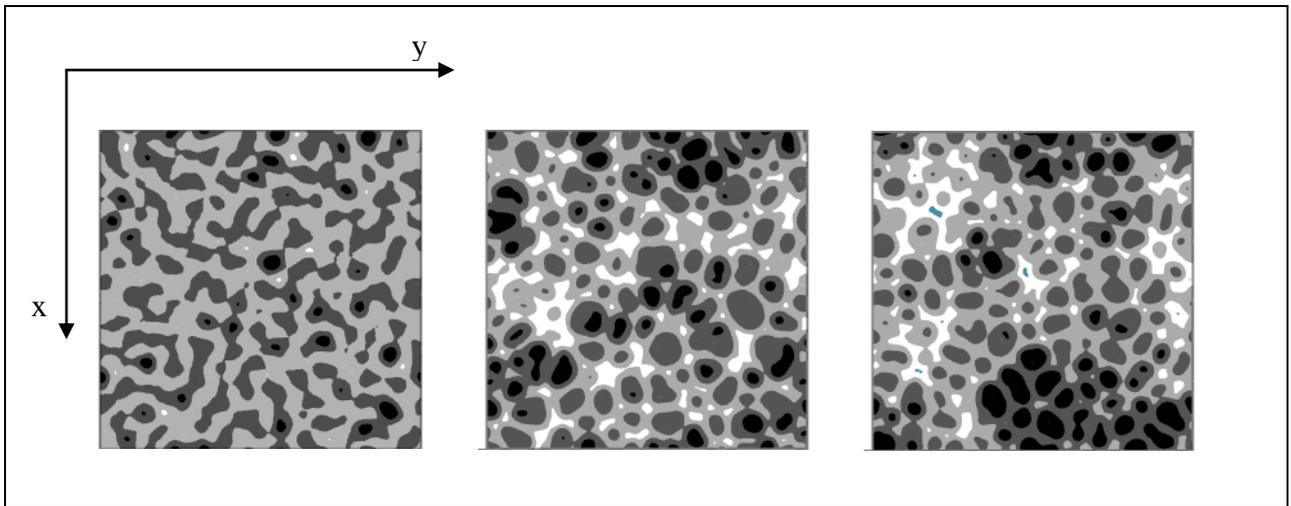

**Fig.3.** Normal incidence (16). Function values $u(x,y)$ in time moments (from left to right) $t = 10^{10}$; $t = 2 \cdot 10^{10}$; $t = 4 \cdot 10^{10}$. Range of values (from left to right) is $-0.2 < u < 0.2$; $-0.5 < u < 0.5$; $-0.6 < u < 0.6$.

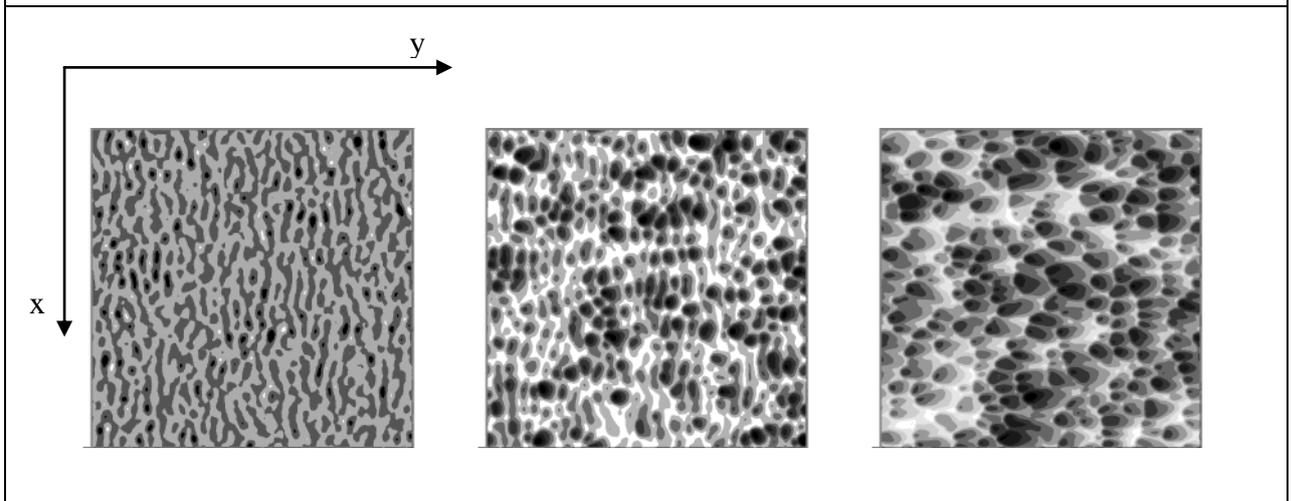

**Fig.4.** Off-normal incidence (17). Function values $u(x,y)$ in time moments (from left to right) $t = 0.5 \cdot 10^9$; $t = 10^9$; $t = 2 \cdot 10^9$. Range of values (from left to right) is $-0.2 < u < 0.2$; $-1.2 < u < 0.6$; $-2.5 < u < 2.5$.

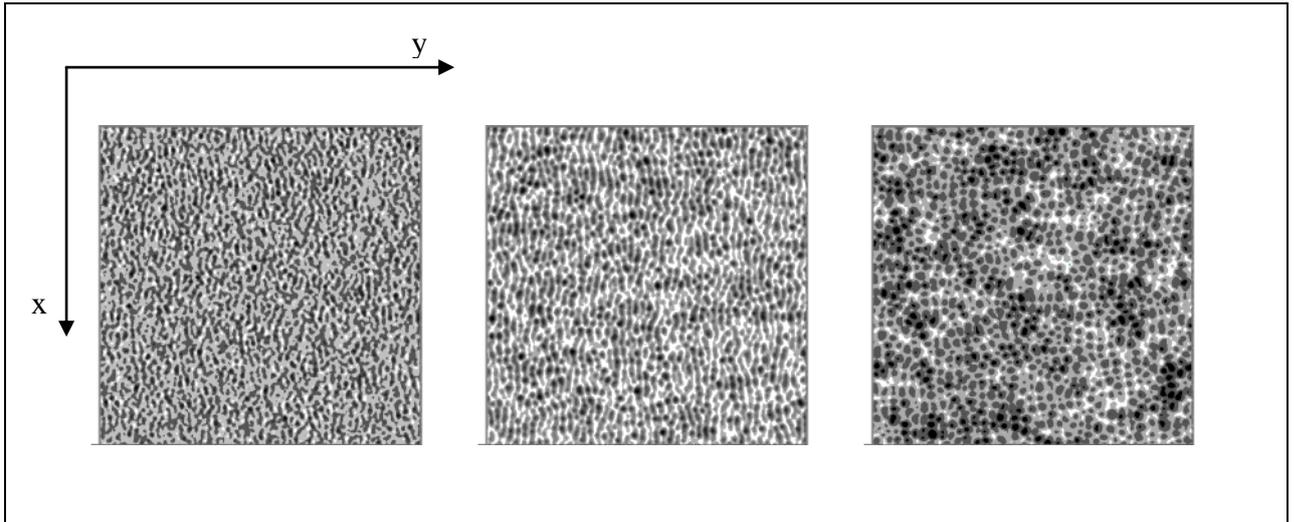

**Fig.5.** Off-normal incidence (18). Function values $u(x,y)$ in time moments (from left to right) $t = 0.2 \cdot 10^8$; $t = 0.5 \cdot 10^8$; $t = 2 \cdot 10^8$. Range of values (from left to right) is $-0.3 < u < 0.2$; $-1.2 < u < 0.6$; $-2 < u < 2$.

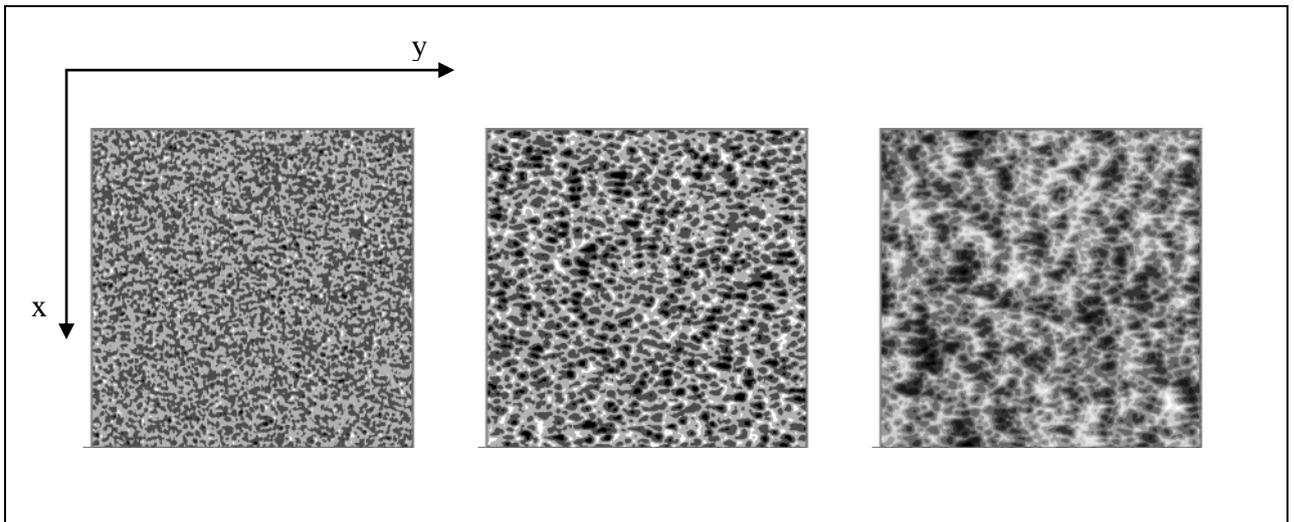

**Fig.6.** Off-normal incidence (19). Function values $u(x,y)$ in time moments (from left to right) $t = 0.2 \cdot 10^8$; $t = 10^8$; $t = 5 \cdot 10^8$. Range of values (from left to right) is $-1 < u < 1$; $-2 < u < 2$; $-4 < u < 4$.

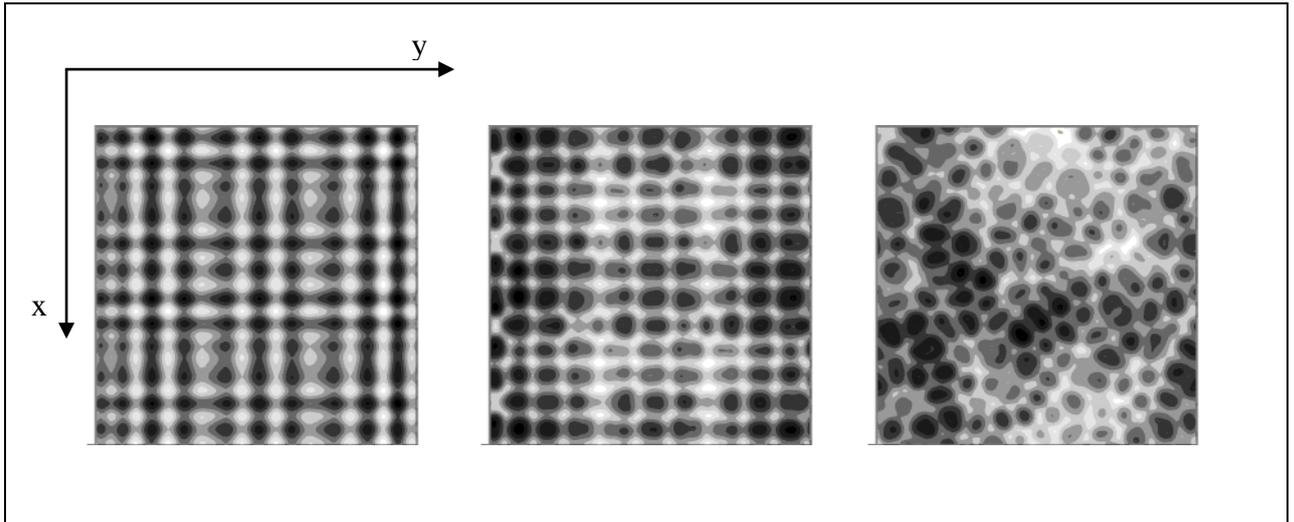

**Fig.7.** Normal incidence (16). Function values $u(x,y)$ in time moments (from left to right) $t = 3 \cdot 10^{10}$; $t = 5 \cdot 10^{10}$; $t = 8 \cdot 10^{10}$. Range of values (from left to right) is $-0.04 < u < 0.04$; $-0.5 < u < 0.5$; $-0.8 < u < 0.8$.

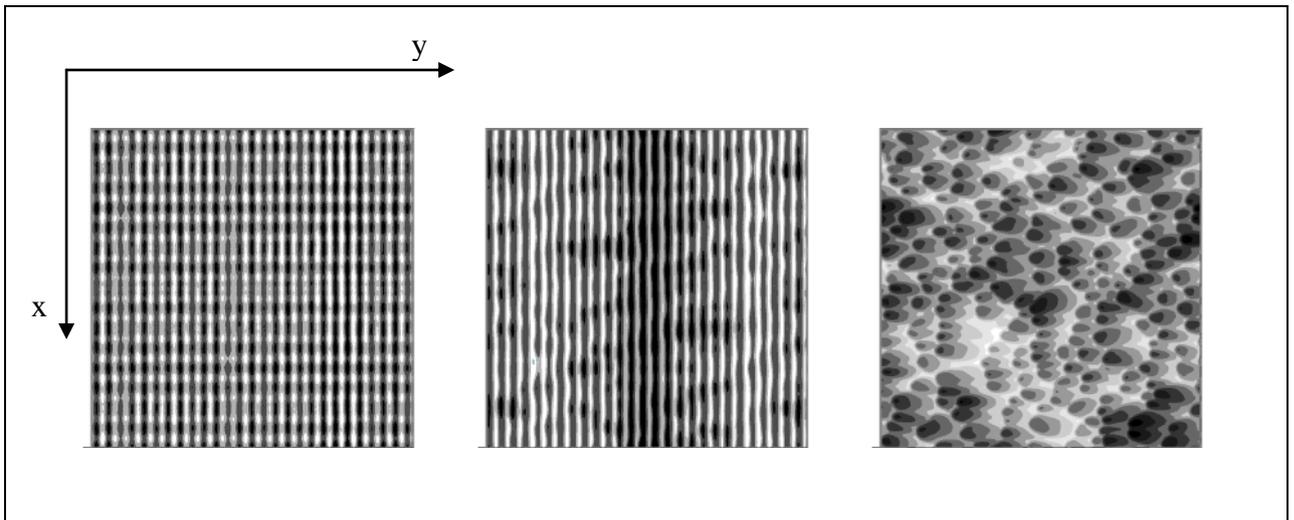

**Fig.8.** Off-normal incidence (17). Function values $u(x,y)$ in time moments (from left to right) $t = 10^9$; $t = 2 \cdot 10^9$; $t = 5 \cdot 10^9$. Range of values (from left to right) is $-0.014 < u < 0.014$; $-0.5 < u < 0.5$; $-3 < u < 3$.

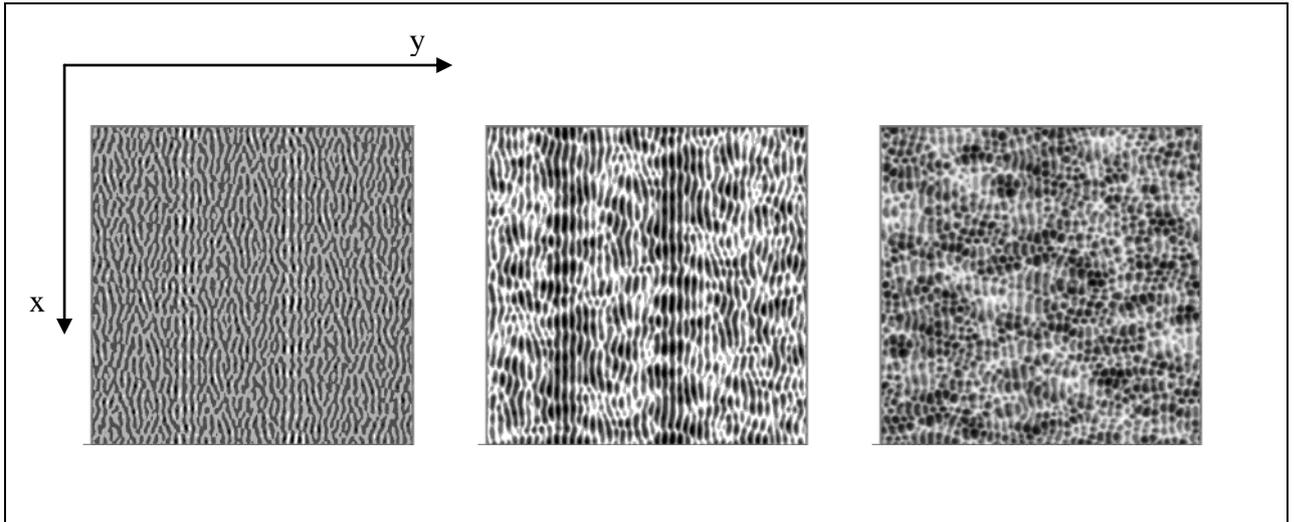

**Fig.9.** Off-normal incidence (18). Function values $u(x,y)$ in time moments (from left to right) $t = 0.5 \cdot 10^8$; $t = 10^8$; $t = 2 \cdot 10^8$. Range of values (from left to right) is $-0.04 < u < 0.04$; $-0.9 < u < 0.6$; $-2 < u < 2$.

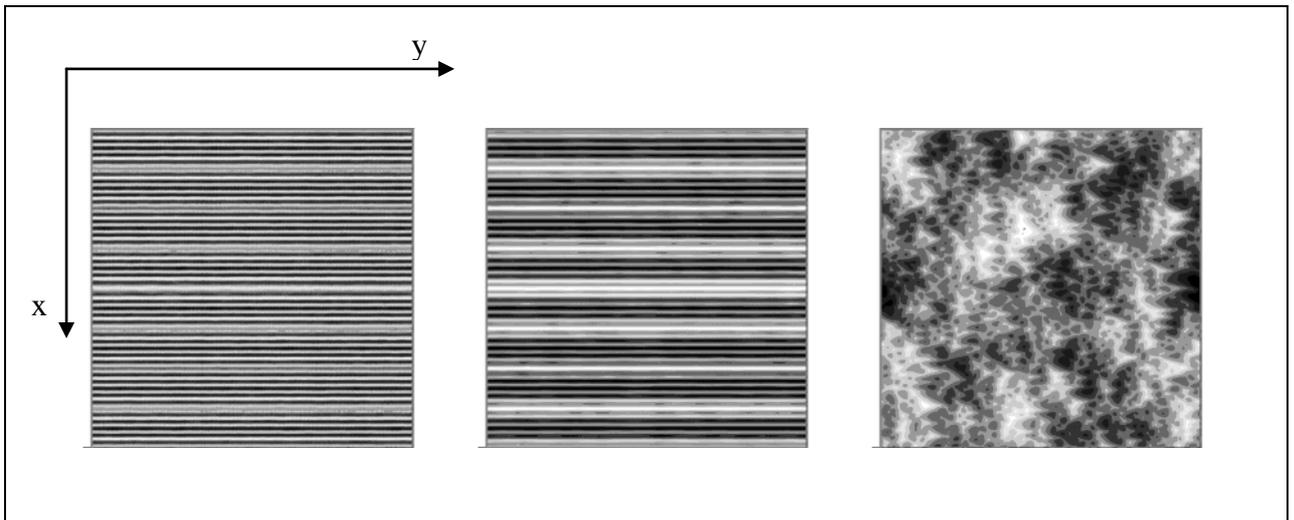

**Fig.10.** Off-normal incidence (19). Function values $u(x,y)$ in time moments (from left to right) $t = 10^8$; $t = 2 \cdot 10^8$; $t = 10^9$. Range of values (from left to right) is $-0.04 < u < 0.04$; $-1.5 < u < 2$; $-6 < u < 6$.

Figs. 8 and 10 show the ripple pattern. At small incidence angles of the plane target surface the pattern is cross ripples (see Fig.8), which wave vector is parallel to the projection of the ion-beam direction on the plane target surface; at gazing incidence the pattern is longitudinal ripples (see Fig.10), which wave vector is perpendicular to the projection of the ion-beam direction on the plane target surface. When the incidence angle is 45 degrees (see Fig.9) the cross ripple less regular than the one shown in Fig.8 is formed. Ripple crest height grows quickly with time and attains a certain maximum value. Then secondary structure of larger scale depressions develops against the ripple pattern and gradually leads to its destruction. At normal incidence (see Fig.7) rectangular lattice made of hillocks and depressions is formed first, then depressions start growing and rectangular symmetry is broken, depressions persist without clear local configuration.

When simulating ion coordinates by sequence of independent random points (see Figs. 3-6) at normal and gazing incidence of ions on the target plane primary structure is not observed. At off-normal incidence (17) and (18) primary structure is not quite developed and is broken faster than in case of simulating ion coordinates by the Holton sequence.

Fig. 11 shows the function $u(x,y)$ calculation results at normal incidence in case when coordinates of ions in incident flux are following:

$$x_k = H_x(p_k + q_k \operatorname{ctg} 60°),$$

$$y_k = H_y q_k,$$

where $(p_k, q_k)$ – Holton sequence members. In this case points $(x_k, y_k)$ are distributed equally on uniform nonrectangular grid which cells are parallelograms with acute angle equaling to 60°. Fig. 11 demonstrates ordered structure from hillocks and depressions with hexagonal symmetry formation. The depressions start growing and then hexagonal symmetry is broken.

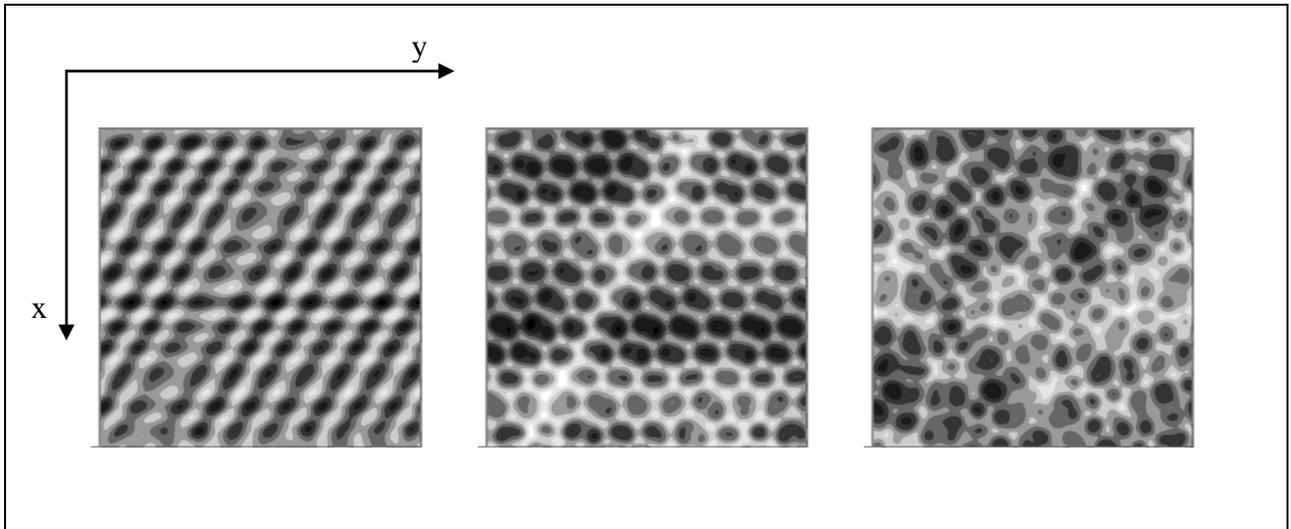

**Fig.11.** Normal incidence (16). Function values $u(x,y)$ in time moments (from left to right) $t = 3 \cdot 10^{10};\ t = 5 \cdot 10^{10};\ t = 8 \cdot 10^{10}$. Range of values (from left to right) is $-0.04 < u < 0.04;\ -0.5 < u < 0.5;\ -0.8 < u < 0.8$.

Figs. 12-15 show graphs of target sputtering depth dependence on time $h_0(t)$. Results got when simulating ion coordinates by sequence of independent random points are marked by a cross; results got when simulating ion coordinates by the Holton sequence are marked by a square; a full line corresponds to solution (7).

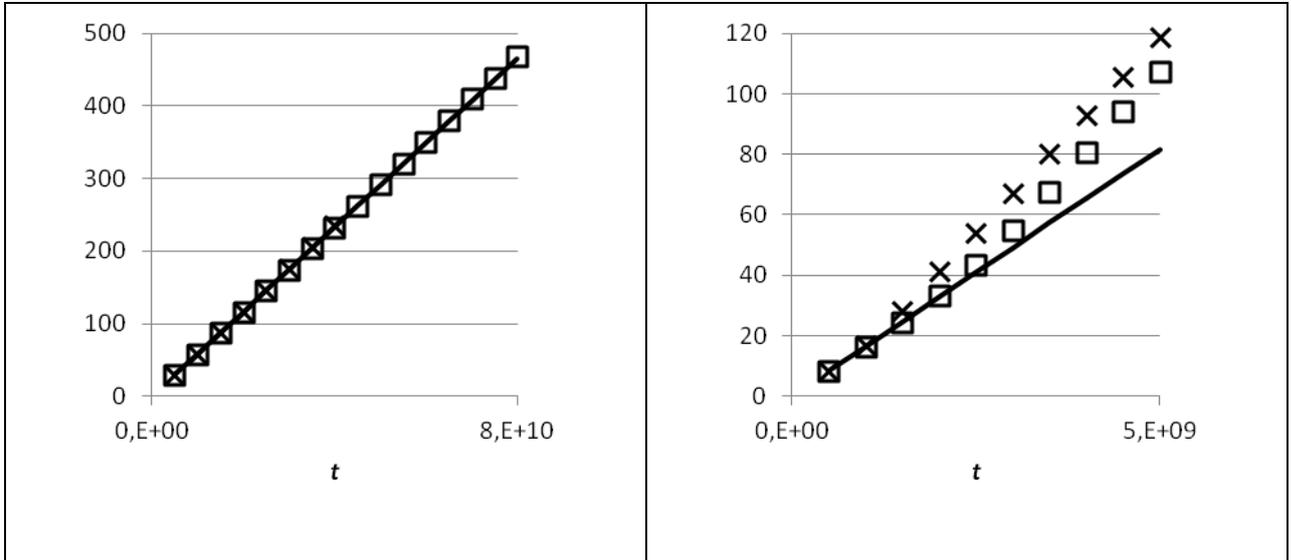

**Fig. 12.** Target sputtering depth dependence on time $h_0(t)$ at normal incidence (16).

**Fig. 13.** Target sputtering depth dependence on time $h_0(t)$ at off-normal incidence (17).

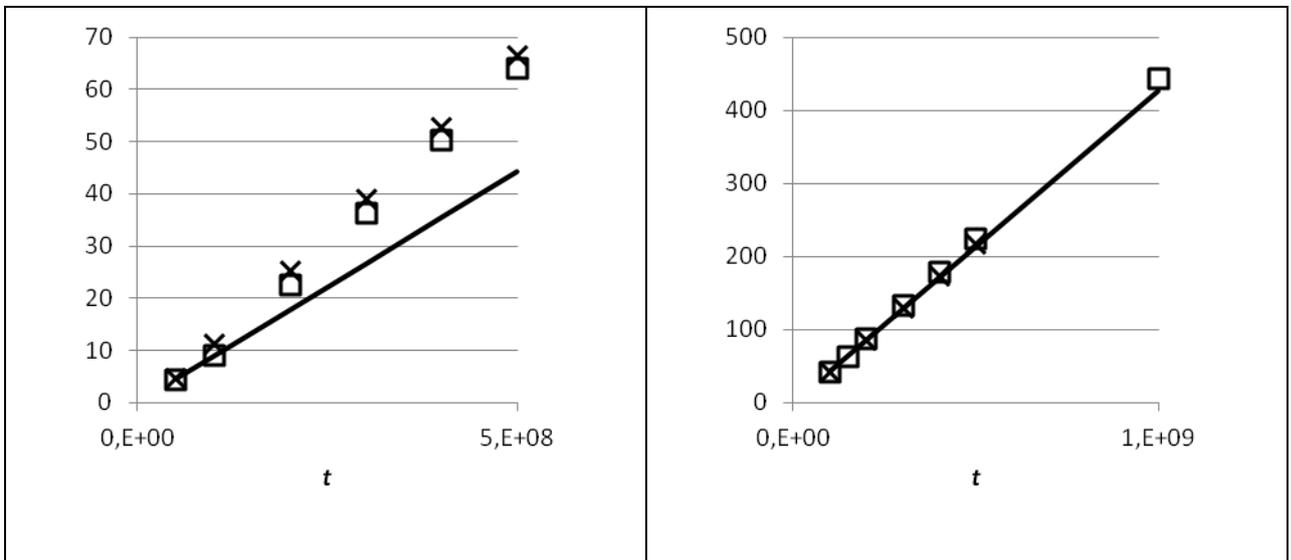

**Fig. 14.** Target sputtering depth dependence on time $h_0(t)$ at off-normal incidence (18).

**Fig. 15.** Target sputtering depth dependence on time $h_0(t)$ at off-normal incidence (19).

Significant deviation of simulation results from the straight line corresponding to the plane target erosion (7) is observed only at off-normal incidence (17) and (18) (see Figs. 13-14) while breaking the cross ripple and forming the pattern secondary structure on the target surface.

## 4. Conclusion

Thus, statistical simulation of pattern formation on plane target surfaces undergoing ion-beam sputtering is shown. In this simulation we use the nonlinear evolutionary equation (3). The difference between this equation and the continuum one (1) is that the ion flux in the integrals (2), (4) is not smooth function. Instead, it is a set of separate incident ions (8).

The simulation with using independent random points of arrival for the incident ions leads to results uncorrelated with continuum model at early times. The continuum model predicts two regimes of ripple formation. The wave vector of the ripples is oriented along the $y$ direction for small enough angles of incidence and along the $x$ direction otherwise. When simulating ion coordinates by sequence of independent random points the ripples are not quite developed or observed.

The simulation with using equally distributed sequence (Holton sequence) of points of arrival for the incident ions leads to results well correlated with continuum model. Both regimes of ripple formation are observed. Moreover, at normal incidence ordered structures from hillocks and depressions are observed. However, we have not physical bases for using Holton sequence. That's why we do not consider the results obtained with using Holton sequence as real physical results. In other works [11-12] using discrete models of pattern formation by ion beam sputtering the authors did not consider the possibility of using equally distributed sequences of points of arrival for the incident ions. Perhaps, it is not quite right. The ion beam sputtering is a controlled process. The incident ions arrival is not absolutely random and the ion flux can contain a deterministic term.

### Acknowledgment

This work was supported by Russian Science Foundation, project to support research carried out by individual research groups No. 14-11-00258.